\documentclass[lettersize,journal]{IEEEtran}
\usepackage{algorithmic}
\usepackage{array}
\usepackage[caption=false,font=footnotesize]{subfig}

\usepackage{textcomp}
\usepackage{stfloats}
\usepackage{url}
\usepackage{verbatim}
\usepackage{graphicx}

\usepackage{balance}

\usepackage{cite}
\usepackage{amsmath,amssymb,amsfonts}
\usepackage{textcomp}
\usepackage{xcolor}
\usepackage{comment}
\usepackage{listings}

\usepackage{soul}
\usepackage[ruled,vlined]{algorithm2e}
\usepackage{subcaption}
\usepackage{url}
\usepackage{makecell}
\usepackage{amssymb}
\usepackage{pifont}
\usepackage{graphicx}

\def\BibTeX{{\rm B\kern-.05em{\sc i\kern-.025em b}\kern-.08em
    T\kern-.1667em\lower.7ex\hbox{E}\kern-.125emX}}
    
\begin{document}

\title{An Online Fragmentation-Aware Scheduler for Managing GPU-Sharing Workloads on Multi-Instance GPUs}
\makeatletter
\newcommand{\linebreakand}{%
  \end{@IEEEauthorhalign}
  \hfill\mbox{}\par
  \mbox{}\hfill\begin{@IEEEauthorhalign}
}
\makeatother

\author{Hsu-Tzu~Ting, Jerry~Chou, Ming-Hung~Chen, and I-Hsin~Chung%
\thanks{Hsu-Tzu~Ting and Jerry~Chou are with the Department of Computer Science, 
National Tsing Hua University, Hsinchu, Taiwan 
(e-mail: hsutzu.ting@lsalab.cs.nthu.edu.tw; jchou@lsalab.cs.nthu.edu.tw).}%
\thanks{Ming-Hung~Chen and I-Hsin~Chung are with IBM T. J. Watson Research Center, 
Yorktown Heights, NY, USA 
(e-mail: minghungchen@ibm.com; ihchung@ibm.com).}%
}

\maketitle

\begin{abstract}
Modern GPU workloads increasingly demand efficient resource sharing, as many jobs do not require the full capacity of a GPU. Among sharing techniques, NVIDIA’s Multi-Instance GPU (MIG) offers strong resource isolation by enabling hardware-level GPU partitioning. However, leveraging MIG effectively introduces new challenges. First, resource contention persists due to shared components such as PCIe bandwidth. Second, GPU fragmentation becomes a critical issue, which is different from prior fine-grained GPU sharing work due to MIG’s limited number of valid MIG configurations. Fragmentation arises not only from spatial discontinuity but also from rigid profile placement constraints, especially after job arrivals and terminations. To address these issues, we propose an online scheduling framework that integrates conditional load balancing, dynamic partitioning, and job migration. 
Our approach dynamically adapts job placement to minimize contention and reorganizes GPU allocations to combat both internal and external fragmentation. Experimental results show that our method significantly improves system efficiency. When all techniques are applied, the makespan improves by up to 35\%.

\end{abstract}


\section{Introduction}
\label{sec:intro}
Today, general-purpose Graphics Processing Units (GPUs) have become one of the most common and power-efficient solutions to provide computing power to drive the development of emerging technologies, such as artificial intelligence (AI), and many other science application domains~\cite{autodrive,real-time-image,nlp,lammps}. Through massive parallel computations, the computing capacity of GPUs has increased significantly by an order of magnitude over the past decade from less than 9 TFLOPS of FP32 performance of NVIDIA TESLA K80 to more than 80 TFLOPS of FP32 performance of NVIDIA B200 GPUs. But at the same time, the price and power consumption of a single GPU device also reach a staggering number. For example, a standalone B200 SXM module costs approximately \$40,000 USD and draws up to 1,000 KW under full load. Hence, maximizing GPU utilization is one of the most critical challenges for data centers and resource managers.

Maximizing GPU utilization is not an easy task because applications often cannot fully utilize the entire GPU device due to their computation behaviors or workload demands. First and foremost, GPUs are still used as accelerators in modern computing systems, hence as suggested by the recent studies~\cite{fastflow,pcie,tccl}, their usage can be bounded by other computing components, such as CPU performance and interconnect memory bandwidth, etc. Secondly, GPUs cannot reach their peak performance without sufficient parallel workloads. With GPUs increasingly used for serving requests, such as AI inferencing, GPUs can easily be underutilized due to the workload variation or user behaviors~\cite{elasticbatch,kubecomp,paris}. Third, as GPUs become more expensive and power-consuming, users may prefer to limit their GPU usage for cost saving. For example, users only need to satisfy their service level agreements (SLAs) in cloud or other pay-as-you-used environments, and jobs only need to be finished before their deadlines in processing systems~\cite{parvagpu,clockwise}.  Last but not least, many techniques are available or under developed for reducing GPU resource demands. Take the popular language model workload as an example. Small language models (SLMs)~\cite{slm} are gaining attention as a more resource-efficient alternative to large language models (LLMs). Techniques, such as Knowledge distillation~\cite{lamini-lm,distillm,gemma2,distill} and Mixture-of-Experts~\cite{Edgemoe,deepseekmoe}, have already been used to achieve competitive performance with reduced resource usage. As a result, it has become common to have GPU-sharing workloads in today's data centers where jobs only demand partial GPU allocations and can be run on a single device for maximizing GPU utilizations.

To support GPU-sharing workloads, many techniques have been proposed to enable resource partitioning and concurrent job execution on GPUs, such as rCUDA~\cite{rcuda}, MPS~\cite{mps}, KubeShare~\cite{kubeshare}, and Multi-Instance GPU (MIG)~\cite{mig}. Among them, MIG stands out for its strong resource isolation.
MIG partitions a single physical GPU into multiple fully isolated instances, referred to as \textit{MIG instances}, each with dedicated compute and memory resources.
It supports a fixed set of partitioning configurations, allowing a GPU to be divided into up to seven MIG instances.
Additionally, MIG supports dynamic reconfiguration, enabling the creation or destruction of GPU instances at runtime without disrupting co-located workloads. This flexibility makes MIG particularly well-suited for multi-tenant environments, where workload isolation and efficient resource utilization are essential.

To leverage the advantage of MIG in job scheduling, several aspects have to be considered. While MIG provides strong isolation by partitioning compute and memory resources at the bare-metal level, certain components, such as PCIe bandwidth, remain shared across instances. Besides common scheduling objectives like load balancing, fragmentation emerges as a critical challenge in MIG-enabled GPU environments.
Although GPU fragmentation has been extensively studied in the context of MPS, where sharing is more fine-grained~\cite{frag}, these insights do not directly apply to MIG. Prior work highlights that GPU allocation must be ``contiguous", meaning that the allocated GPU resource must reside on the same physical GPU. For example, a job requiring a full GPU cannot be served by aggregating partial GPUs from two different GPUs. 
However, under MIG, even when sufficient resources are available on a single GPU, instance creation may still fail.
This is because MIG supports only a fixed set of partition configurations, and the available resources may not align with any valid configuration. As a result, two GPUs with identical residual resources may differ in availability depending on their configuration state. 
This limitation introduces a new form of fragmentation and inefficiency.

To fully leverage the advantages of MIG, including memory isolation and dynamic partitioning, while addressing key challenges like resource contention and fragmentation, we propose a comprehensive online method that incorporates \textit{conditional load balancing}, \textit{dynamic partitioning}, and \textit{job migration}. Conditional load balancing mitigates resource contention by distributing workloads across GPUs based on a user-defined load-balancing threshold. This strategy prevents the over-concentration of tasks on a single GPU. To address fragmentation, dynamic partitioning and job migration are employed. Dynamic partitioning creates MIG instances that precisely match the requirements of incoming jobs, reducing internal fragmentation. On the other hand, job migration redistributes workloads to optimize the use of residual GPU capacity, thereby reducing external fragmentation. Our main contributions are summarized as follows:

\begin{itemize}
\item We discuss GPU fragmentation specific to MIG environments, bridging a gap left by prior work, which primarily targets fine-grained sharing, such as MPS.
\item We propose a novel online scheduling framework for MIG-enabled GPUs that integrates conditional load balancing, dynamic partitioning, and job migration to improve both resource utilization and job performance.
\item We deliver experimental evaluation, demonstrating that our proposed method reduces workload completion time compared to baseline approaches from 13\% to 35\%.
\end{itemize}

\section{Motivation}
\label{sec:mot}
In this section, we explain the MIG feature in NVIDIA GPUs as well as its advantages and challenges. These insights motivate us to develop a scheduling framework that harnesses MIG’s resource isolation and dynamic partitioning while addressing contention and fragmentation to improve overall GPU efficiency.

\subsection{Multi-Instance GPU (MIG)}
Multi-Instance GPU (MIG) is a hardware feature that allows GPUs (starting with the NVIDIA Ampere architecture) to be partitioned into several smaller GPU instances. It is beneficial for workloads that do not saturate the GPU's computing capacity. This scenario commonly occurs because the computing power of the GPU greatly increases year by year, making it realistic to run multiple GPU workloads in parallel to maximize GPU efficiency.

MIG feature splits GPUs in an isolated manner, with separate streaming multiprocessors, GPU memory, and L1/L2 cache. To create a MIG instance, creating a GPU instance (GI) is required, followed by creating a compute instance (CI). A GI includes one or more compute slices and memory slices, while the naming convention is the combination of them. For example, a 1g.5gb GI indicates that 1 compute slice and 5 GB memory are contained. Although a GI can be subdivided into several CIs, we always create a CI that fully utilizes the parent GPU instance's SM for simplicity. In this paper, we target the MIG instances of A100 40 GB, including 1g.5gb, 2g.10gb, 3g.20gb, and 4g.20gb, and denote them by their compute slices as 1g, 2g, 3g, and 4g respectively. The configuration of GPU is denoted by a list of instances such as [4g,3g].

\subsection{Advantages of MIG}
Compared with other GPU sharing techniques like Multi-Process Service (MPS)~\cite{mps}, MIG provides better isolation. Although MPS partitions the computing resources, memory resources, including caches, and memory control, are not isolated. Sharing memory resources among workloads tends to cause interference issues. In contrast, the MIG feature partitions both computing and memory resources. Each MIG instance has separate memory paths to minimize the interference of concurrent jobs within the GPU.

Another advantage of MIG is the support of dynamic partitioning. Once the GPU is in MIG mode, GIs and CIs can be configured dynamically. A GPU with configuration [4g,3g] can be reconfigured to [4g,2g,1g] by destroying the 3g MIG instance and then creating the 2g and 1g MIG instance, as long as the 3g MIG instance is idle. The reconfigurability provides flexibility in adjusting the GPU partition according to the current workloads.

\subsection{Challenges of MIG}
Although the MIG feature partitions both the computing and memory resources of the GPU, contention still occurs in certain circumstances. First, the PCIe bandwidth is shared among co-running workloads. If one workload consumes a significant amount of PCIe bandwidth, it can interfere with others, leading to performance degradation~\cite{pcie}. Second, a study reveals that the MIG feature does not partition the last-level TLB and leaves it shared~\cite{tlb}. The co-running workloads can cause severe TLB thrashing, which also degrades performance.

Another challenge is the inflexibility of MIG configuration. The number of valid GI profiles is limited. For example, A100 40 GB can be partitioned into 1g, 2g, 3g, 4g, or 7g, but not 5g and 6g. The number of valid GPU configurations is also limited because each GI is allowed to be created on certain indexes, as shown in Table~\ref{tab:profile}. This constraint easily causes fragmentation, which is explained in Section~\ref{sec:frag}.

\begin{table}[h]
    \centering
    \renewcommand{\arraystretch}{1.2} 
    \begin{tabular}{|c|c|c|c|c|}
        \hline
        \textbf{Profile} & \textbf{Compute Slices} & \textbf{Memory Slices} & \textbf{Starting Idx} & \textbf{Size}\\ \hline
        7g.40gb    & 7 & 8 & 0 & 8 \\ \hline
        4g.20gb    & 4 & 4 & 0 & 4 \\ \hline
        3g.20gb    & 3 & 4 & 0,4 & 4 \\ \hline
        2g.10gb    & 2 & 2 & 0,2,4 & 2 \\ \hline
        1g.10gb    & 1 & 2 & 0,2,4,6 & 2 \\ \hline
        1g.5gb    & 1 & 1 & 0,1,2,3,4,5,6 & 1 \\ \hline
    \end{tabular}
    \caption{GPU Instance Profiles on A100 40GB}
    \label{tab:profile}
\end{table}

\section{Fragmentation}
\label{sec:frag}
While GPU fragmentation has been widely discussed~\cite{frag}, the definition in the context of MIG remains unclear. Previous works highlight the contiguity of GPU allocation. For example, when a job requests one full GPU, two partial GPUs (e.g., 0.4 GPU + 0.6 GPU) fall into fragments because the available GPU resources are separated on two different GPUs rather than contiguous on one single GPU.
However, in the context of MIG, discontinuity is not the only cause of fragmentation. The NVIDIA MIG feature partitions GPUs into multiple small GPU slices, but the number of supported profiles is limited. In this section, we elaborate on how the MIG constraint influences GPU fragmentation and explain two forms of fragmentation in MIG environments: \textit{external} and \textit{internal} fragmentation.

\subsection{External Fragmentation}
The traditional definition of external fragmentation in GPUs refers to partially available GPU resources that are insufficient to accommodate incoming jobs. These fragmented GPU segments are often scattered across the cluster and cannot be aggregated to serve larger workloads. Consequently, the continuity of unallocated GPU resources is considered essential to avoid fragmentation. However, in MIG-enabled GPU environments, discontinuity is not the only cause of external fragmentation.
Due to the limitation of the GPU instance profile, GPU instances can only be created on certain indexes. That is, even the contiguous GPU resource does not guarantee that a MIG instance can be created. For example, in Figure~\ref{fig:ext_frag}, GPU 1 has a contiguous 4g space, but a 4g MIG instance cannot be created because it can only be created at index 0. Although the available partial GPUs are continuous, external fragmentation still exists because of the limited number of valid configurations.

\begin{figure}[tp]
    \centering
    \includegraphics[width=1\linewidth]{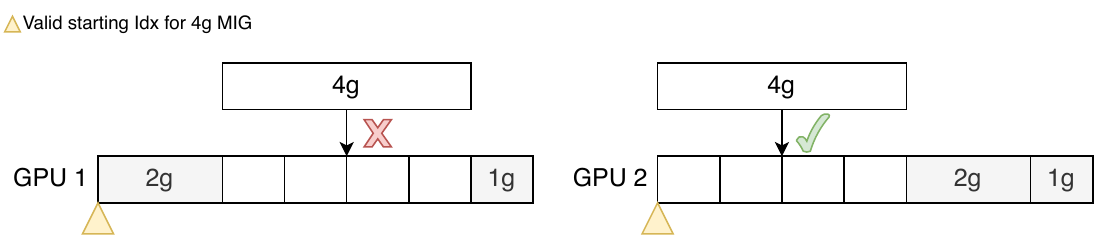}
    \caption{GPU 1 has a continuous 4g space but cannot serve a 4g MIG instance due to external fragmentation.}
    \label{fig:ext_frag}
\end{figure}

The external fragmentation indicates that partial GPUs with the same amount of remaining GPU resources may not be treated identically. GPU 1 and GPU 2 in Fig.~\ref{fig:ext_frag} have the same remaining GPU resources but differ in \textit{availability}. A 4g MIG instance can be created on GPU 2 but not on GPU 1. This setting is fundamentally different from previous work on GPU fragmentation, where GPUs with the same remaining resources were considered the same. 

External fragmentation occurs when the placement of GPU instances is arbitrary and when the job departs. We empirically discovered that NVIDIA creates GPU instances in a way that minimizes external fragmentation and maximizes GPU availability. When creating a 2g MIG instance, rather than creating on index 0 or 2, it prefers index 4 because occupying indexes 0-4 restricts the GPU's availability to serve a 4g MIG instance. However, external fragmentation still easily occurs after job termination. In Fig.~\ref{fig:ext_frag_job_leaves}, the initial placement of GPU instances is properly organized. However, when the short-term jobs are finished, the continuous space becomes external fragmentation. Our work addresses this situation through intra-GPU migration, which reorganizes the job placement to maximize the GPUs' availability.

\begin{figure}[tp]
    \centering
    \includegraphics[width=1\linewidth]{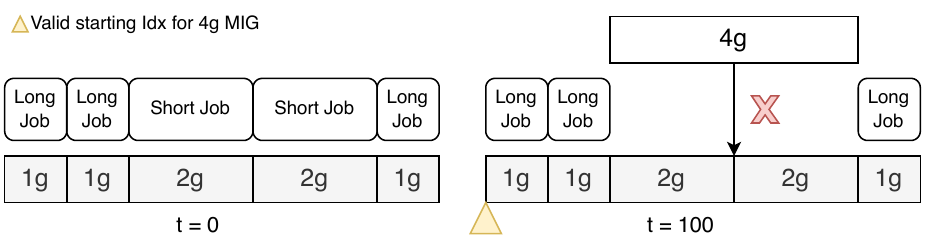}
    \caption{External fragment occurs when jobs are finished.}
    \label{fig:ext_frag_job_leaves}
\end{figure}

\subsection{Internal Fragmentation}
Internal fragmentation traditionally refers to the condition where the memory allocated to a process exceeds the memory actually requested or utilized by that process. In MIG-enabled GPU environments, a MIG instance is created from the partial GPUs and then assigned to jobs. In this context, we define internal fragmentation as a situation where the MIG instance assigned to the job exceeds the actual requirements of the job. The surplus GPU capacity within the MIG instance cannot be reallocated to other jobs, resulting in underutilization of resources.
Some orchestrators, such as Kubernetes, explicitly prohibit GPU over-allocation to prevent internal fragmentation. When the Kubernetes scheduler selects a node for a job, the node must contain a MIG instance that exactly matches the job’s GPU resource request. A node is not eligible if it hosts a MIG instance that is either larger or smaller than the requested size. 
This design is understandable in terms of reducing the waste of GPU resources. In our case, we also prohibit internal fragmentation and dynamically partition a GPU into the exact instance that a job requests.
\section{Design \& Implementation}
In addition to common scheduling objectives such as load balancing, a scheduler operating in a MIG-enabled environment must also account for fragmentation to fully utilize GPU resources. As discussed in the previous section, MIG introduces unique fragmentation challenges due to its fixed partitioning profiles and placement constraints. In this section, we present the design and implementation of the online fragmentation-aware scheduler that addresses these challenges through conditional load balancing, dynamic partitioning, and job migration.

\subsection{Problem Definition} 
Our goal is to design a system that leverages the advantages of MIG while simultaneously addressing contention and fragmentation. We focus on the single-node scenario as the broader node-level scheduling decisions such as selecting which node a job should run on, are orthogonal to our design. Once the node-level scheduler selects a node, our GPU-level scheduler determines the appropriate GPU for each job. To tackle these challenges, we divide the problem into two phases: scheduling upon job arrival and migration upon job departure.

Importantly, we assume that each incoming job specifies a fixed MIG instance size, similar to the Kubernetes setting where each pod needs to declare its required resources. Automatically selecting the most suitable instance size is a complex scheduling problem that involves predicting job performance, and this has been studied extensively in prior work~\cite{mig-serving,elasticbatch,parvagpu,migrator,paris}. We view this problem as complementary but orthogonal to our goals. Therefore, we explicitly consider MIG instance sizes as predetermined input and focus on improving scheduling under the fixed-size constraints. 
\begin{figure}[tp]
    \centering
    \includegraphics[width=0.99\linewidth]{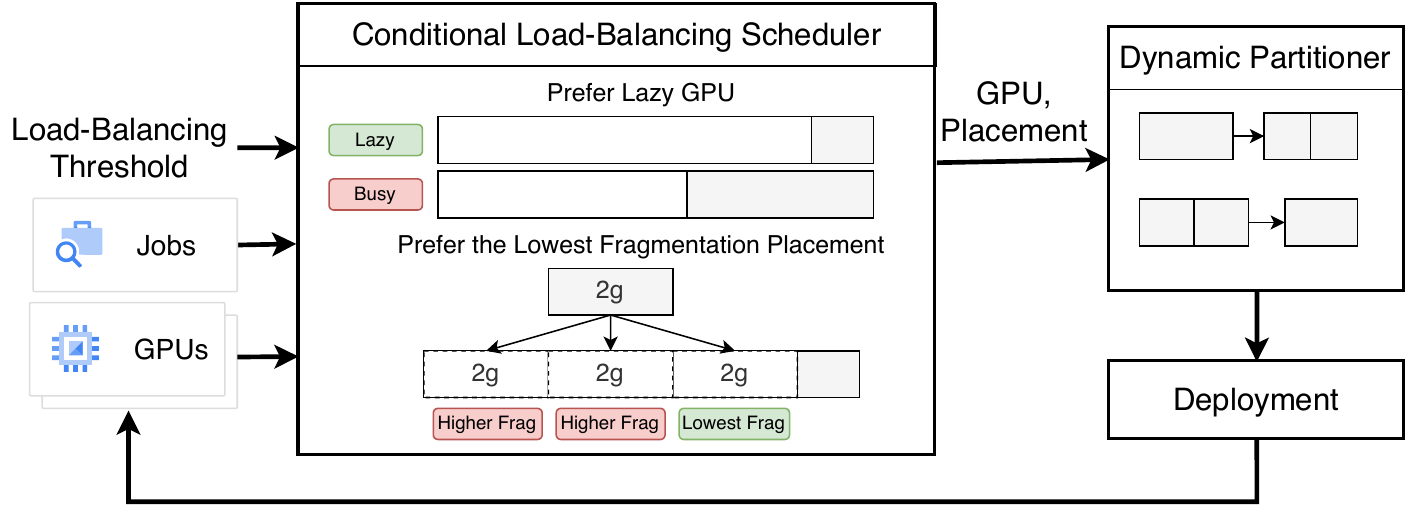}
    \caption{Scheduling upon job arrival. The conditional load-balancing scheduler prefers the Lazy GPU (the GPU with utilization lower than the load-balancing threshold) and the placement with the least fragmentation.}
    \label{fig:mig-sched-prob}
\end{figure}

\begin{figure}[tp]
    \centering
    \includegraphics[width=0.99\linewidth]{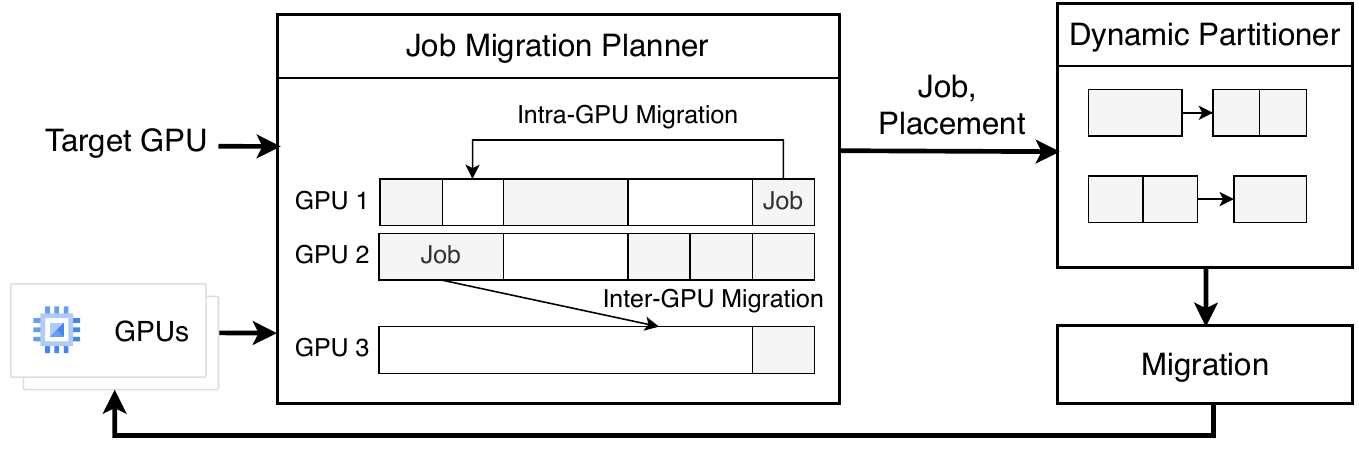}
    \caption{Migration upon job departure from target GPU. The job migration planner formulates an intra-GPU migration or an inter-GPU migration plan to balance the workload across GPUs and minimize GPU fragmentation.}
    \label{fig:mig-migrate-prob}
\end{figure}

\textbf{Job Arrival Scheduling Problem.} When a job arrives, the scheduler has to determine an appropriate MIG instance for execution as shown in Fig.~\ref{fig:mig-sched-prob}. The decision impacts the severity of contention and fragmentation in the system. Given a job \( J \) with the requested MIG instance profile \( M \), the scheduler determines the pair \( (G_i, P) \), where \(G_i\) is the GPU that job \(J\) is going to run on, and \(P = (st, sz) \) is the MIG placement. \( st \) denotes the starting index, and \( sz \) denotes the size of the MIG instance. 

The scheduler determines \( (G_i, P) \) based on the states of all GPUs, denoted as \( \mathcal{G} = \{ G_1, G_2, \dots, G_n \} \). Each GPU consists of compute slices and memory slices. Each compute slice and memory slice is exclusively allocated to a single MIG instance, and each MIG instance is exclusively assigned to one job. 
Therefore, \( (G_i, P) \) must satisfy two conditions: 
\begin{align}
    & \text{(1) Validity:} \quad Valid(M, P) = 1 \label{eq:validity} \\
    & \text{(2) Availability:} \quad Avail(G_i, P) = 1 \label{eq:availability}
\end{align}
 \(Valid(M, P)\) and \(Avail(G_i, P)\) are indicator functions.  \(Valid(M, P)\) returns 1 if the placement \( P \) is one of the valid profile for \( M \) as shown in Table~\ref{tab:profile}. \(Avail(G_i, P)\) returns 1 if both compute slices and memory slices at placement \(P\) are not yet occupied by any job on GPU \(G_i\). 

\textbf{Job Departure Migration Problem.}
When a job \( J \) completes execution and departs from GPU \(G_i\), the scheduler may initiate a migration process to balance the GPU workload and reduce fragmentation as shown in Fig.~\ref{fig:mig-migrate-prob}. Considering the states of all GPUs \( \mathcal{G} = \{ G_1, G_2, \dots, G_n \} \), the set of currently running jobs \( \mathcal{J} = \{ J_0, J_1, \dots, J_m \} \), and their respective placements \( \mathcal{P} = \{ P_0, P_1, \dots, P_m \} \), the scheduler determines whether a job migration is required. If migration is necessary, the scheduler selects \( J_k \) and determines a new placement \(P_k'\) for it on \(G_i\). The migration process iterates until no further jobs require relocation. To ensure a non-disruptive migration, a replica of the migrating job is created on the target GPU before its execution is terminated on the original GPU.

\subsection{Fragmentation Measurement}
We define the fragmentation cost function \( FragCost(G_i) \) that quantifies the severity of fragmentation on a given GPU \(G_i\) by capturing the average unavailability of each MIG instance profile due to fragmentation. 

Let \( \mathcal{M} = \{ M_1, M_2, \dots, M_k \} \) be the set of available MIG instance profiles (such as 1g.5gb, 2g.10gb, etc.), where each \( M_j \) requires a contiguous allocation of compute slices \(cs_j\) and memory slices \(ms_j\). They can only be created at specific indices as shown in Table~\ref{tab:profile}. The ideal number of \(M_j\) that can be created on \(G_i\) is 
\begin{equation}
ideal\_mig\_num(G_i, M_j) = min(\lfloor RC_i / cs_j \rfloor, \lfloor RM_i / ms_j \rfloor)
\label{ideal}
\end{equation}

where \(RC_i\) is the remaining compute slice and \(RM_i\) is the remaining memory slice of \(G_i\). This equation does not consider MIG constraints. 

If MIG constraints are considered, the number of valid placements for \(M_j\) on \(G_i\) is given by:
\begin{multline}
{feasible\_mig\_num}(G_i, M_j) =
\Big| \big\{ P \mid 
Valid(M_j, P) \wedge {} \\
Avail(G_i, P) \big\} \Big|
\label{feasible}
\end{multline}
As indicated by the equation, a feasible MIG placement should be valid with respect to the MIG profile and should be available with respect to the GPU.

Using \(ideal\_mig\_num\) and \(feasible\_mig\_num\), we define the fragmentation cost function as :

\begin{equation}
FragCost(G_i) = 1 - \frac{1}{|\mathcal{M}|} \sum_{j=1}^{|\mathcal{M}|}\frac{feasible\_mig\_num(Gi, Mj)}{ideal\_mig\_num(Gi, Mj)}
\label{cost_function}
\end{equation}
which represents the average unavailability of MIG instances on \(G_i\).

\subsection{Job Arrival Scheduling Method}
We propose a conditional load-balancing and fragmentation-aware scheduling method tailored for MIG-enabled GPUs, designed to balance execution performance and GPU availability. When a job \( J \) arrives, the scheduler has to select a pair \( (G_i, P) \), where \(G_i\) and \(P\) are the selected GPU and MIG placement. The decision navigates the inherent trade-off between minimizing the job’s execution time, which is affected by GPU contention, and reducing GPU fragmentation, which impacts the GPU availability.

The core innovation of our approach lies in a threshold-based conditional load-balancing mechanism. Guided by a user-defined parameter \(t\), the scheduler classifies each GPU as either ``Lazy" if its current utilization is below \(t\), or ``Busy" if its utilization exceeds \(t\). The load-balancing threshold dynamically controls the aggressiveness of job spreading because the scheduler prefers the Lazy GPU over the Busy GPU. Lower \(t\) causes the scheduler to spread jobs early, while a higher \(t\) prioritizes fragmentation minimization and condenses the jobs on the same GPU.

In addition to load balancing for GPU selection, we incorporate fragmentation-aware MIG placement and the partitioning reuse strategy. For each valid MIG placement candidate on a selected GPU, the scheduler evaluates the fragmentation cost \(FragCost\) by hypothetically applying the placement and computing its impact on the GPU’s future configurability. The placement that yields the lowest fragmentation cost is preferred. To further reduce scheduling overhead, we prioritize the reuse of existing MIG instances when multiple placements produce the same fragmentation cost. This optimization avoids unnecessary re-partitioning, thereby improving responsiveness and reducing reconfiguration time.

The detailed steps of the job placement are illustrated as follows:

\begin{itemize}
\item Step 1: Classify each GPU as Lazy or Busy based on the user-defined load-balancing threshold \(t\).
\item Step 2: For Lazy GPUs, enumerate all feasible MIG placements. For each placement, compute the resulting fragmentation using the fragmentation cost function, and select the placement with the minimum fragmentation.
\item Step 3: Among placements with equal fragmentation cost, prefer those that reuse existing MIG partitions to minimize reconfiguration overhead.
\item Step 4:  If no feasible placement is found on Lazy GPUs, repeat Steps 2 and 3 for Busy GPUs.
\item Step 5: If no feasible placement is found on Busy GPUs, queue the job and schedule it in first-come-first-served (FCFS) order.
\end{itemize}

If the selected pair \((G_i, P)\) corresponds to an existing MIG partition, the job is scheduled directly. Otherwise, the GPU is dynamically partitioned to create the required instance before dispatching the job.

\subsection{Job Departure Migration Method}
To maintain GPU availability by minimizing fragmentation, we introduce a fragmentation-aware job migration strategy triggered upon job completion. When a Job \(J\) completes and departs from GPU \(G_i\), the scheduler determines whether to migrate jobs to balance the GPU workload and reduce fragmentation. There are two forms of migration, \textit{intra-GPU migration} and \textit{inter-GPU migration}. The decision is guided by the updated status of \(G_i\). If GPU \(G_i\) remains Busy after the job's departure, the scheduler considers intra-GPU migration to minimize fragmentation within \(G_i\). Conversely, if GPU \(G_i\) is in the Lazy status, then the scheduler considers inter-GPU migration to balance the workload across GPUs while also considering fragmentation reduction.

If \(G_i\) is classified as Busy after a job departure, the scheduler aims to reorganize the job placement on \(G_i\) to achieve a more compact and reusable MIG configuration. The process is as follows:

\begin{itemize}
    \item Step 1: Enumerate all condidate migrations \((J, P)\), where \(J\) is a running job on \(G_i\), and \(P\) is a valid and unoccupied placement for \(J\) on the same GPU. For each candidate, calculate the resulting fragmentation of \(G_i\) by assuming job \(J\) is migrated to \(P\).
    \item Step 2: Select the pair \((J, P)\) which minimizes the fragmentation cost on \(G_i\) and migrate \(J\) to \(P\) accordingly.
    \item Step 3: Repeat Steps 1 and 2 until no candidate migration can further reduce the fragmentation of \(G_i\).
\end{itemize}

If the departure of job J causes \(G_i\) to become Lazy, the GPU becomes eligible to receive jobs from other Busy GPUs in order to alleviate load imbalance. The process is as follows:

\begin{itemize}
    \item Step 1: Enumerate running jobs on Busy GPUs. Only consider jobs where migrating from \(G_j\) to \(G_i\) can result in a lower load on \(G_i\) than on \(G_j\).
    \item Step 2: For each eligible job \(J\), calculate the resulting fragmentation on \(G_j\) by assuming \(J\) is migrated to \(G_i\).
    \item Step 3: Select the job \(J\) which minimizes fragmentation on \(G_j\) and migrate it to \(G_i\) using the placement that minimize fragmentation on \(G_i\).
    \item Step 4: Repeat Steps 1 to 3 until no job meets the criteria in Step 1. 
\end{itemize}

To ensure zero downtime during migration, the new MIG instance is created, and the job is scheduled on the target GPU before terminating the original job. This migration strategy ensures uninterrupted service and minimal scheduling overhead.

\subsection{Time Complexity}
As our method operates in an online scheduling setting, analyzing its time complexity is essential to understanding its practical feasibility. This section examines the time complexity of the fragmentation cost function, job arrival scheduling, and job departure migration. 

Let \(g\) denote the number of GPUs, \(m\) denote the number of MIG profiles and \(n\) denote the maximum number of isolated GPU instances that the NVIDIA GPU can support. Both \(m\) and \(n\) are constants defined by the hardware. As shown in Table~\ref{tab:profile}, there are \(m = 6\) different MIG profiles, and the NVIDIA A100 GPU can partition up to \(n=7\) isolated GPU instances.

The time complexity of the \(FragCost(G_i)\) (Equation~\ref{cost_function}) is \(O(mn)\) as it evaluates across \(m\) MIG profiles for GPU while  \(ideal\_mig\_num(G_i, M_j)\) (Equation~\ref{ideal}) and \(feasible\_mig\_num(G_i, M_j)\) (Equation~\ref{feasible}) take \(O(1)\) and \(O(n)\) respectively. Given that \(m\) and \(n\) are constants for a specific GPU model, \(T\_{frag}\) simplifies to \(O(1)\).

Job arrival scheduling evaluates all \(g\) GPUs, classifying each as Lazy or Busy and checking up to \(n\) possible placements. These steps yield a total complexity of \(O(g + gn \cdot T\_{frag}) = O(g)\), since \(T\_{frag}\) and \(n\) are constant. 
Job departure migration includes intra-GPU migration and inter-GPU migration scenarios. 
Intra-GPU migration considers the cost for all possible single-job migrations within the target GPU, resulting in at most \(n^2\) combinations and a complexity of \(O(n^2) \times T\_{frag}\ = O(1)\).
Inter-GPU migration considers up to \(g \cdot n\) jobs migrating to one of \(n\) placements on the target GPU, leading to a complexity of \(O(gn \cdot n\ \cdot T\_{frag}) = O(g)\).

\section{Experimental Evaluation}
\label{sec:exp}
\subsection{Experiment Setup}
\subsubsection{Testbed}
Our testbed consists of a single node, equipped with 4 NVIDIA A100 40GB GPUs. The node runs on Ubuntu 20.04.6 LTS, and the NVIDIA Driver version is 570.86.15. The scheduler is developed in Go, utilizing Go bindings for the NVIDIA Management Library (NVML) to manage the GPU's configuration.

\subsubsection{Workloads}
The workloads contain several tasks to complete, and each task contains inference queries. We use the Poisson distribution recommended by the MLPerf inference benchmark~\cite{mlperf} to model how frequently the tasks are sent to the inference server. Each task requests a MIG instance among 1g.5gb, 2g.10gb, 3g.20gb and 4g.20gb and specifies a model among opt-6.7b, opt-13b, bloom-1b7, and bloom-7b1. We run opt-6.7b and bloom-1b7 on either 1g.5gb or 2g.10gb and run opt-13b and bloom-7b1 on either 3g.20gb or 4g.20gb. The MIG memory is limited, so some parameters are offloaded to the CPU memory. The request and response tokens of the query are selected from BurstGPT~\cite{burstgpt}, excluding outliers. We generate four workloads with different query characteristics and arrival rates as shown in Table~\ref{tab:mig-workloads}. ``Normal" queries are randomly selected from BurstGPT, while ``Long" queries are randomly selected from the top 50\% in length.

\subsubsection{Load-Balancing Threshold} The threshold is set to 0.4 unless explicitly specified. This value ensures that jobs are packed before a GPU becomes half-loaded.

\begin{table*}[!t]
    \centering
    \renewcommand{\arraystretch}{1.1} 
    \begin{tabular}{|c|c|c|} \hline 
 Workload & Mean Arrival Time & Query Type\\ \hline 
         Normal(25)&  25& Randomly Selected\\ \hline  
         Long(25)&  25& Randomly Selected from Top 50\% in Length\\ \hline  
         Normal(50)&  50& Randomly Selected\\ \hline  
         Long(50)&  50& Randomly Selected from Top 50\% in Length\\ \hline 
         \end{tabular}
    \caption{Four workloads used to evaluate.}
    \label{tab:mig-workloads}
\end{table*}

\begin{figure*}[t]
  \centering
  \subfloat[Bloom1b7]{%
    \includegraphics[width=0.24\linewidth]{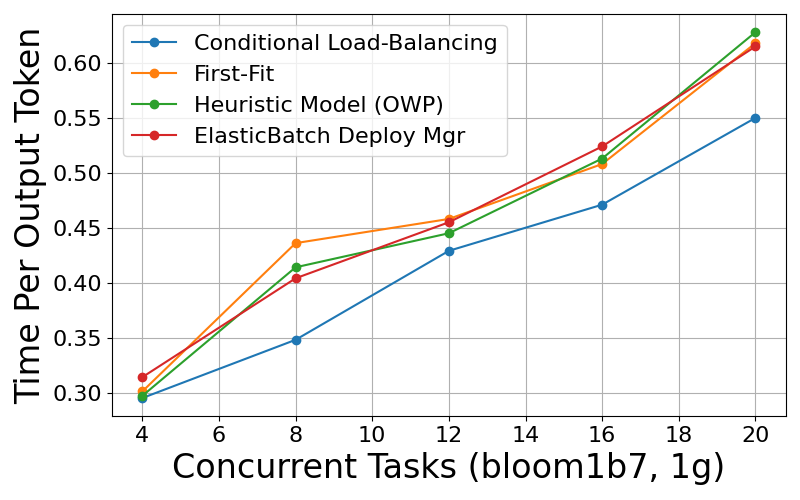}}
  \hfill
  \subfloat[Bloom7b1]{%
    \includegraphics[width=0.24\linewidth]{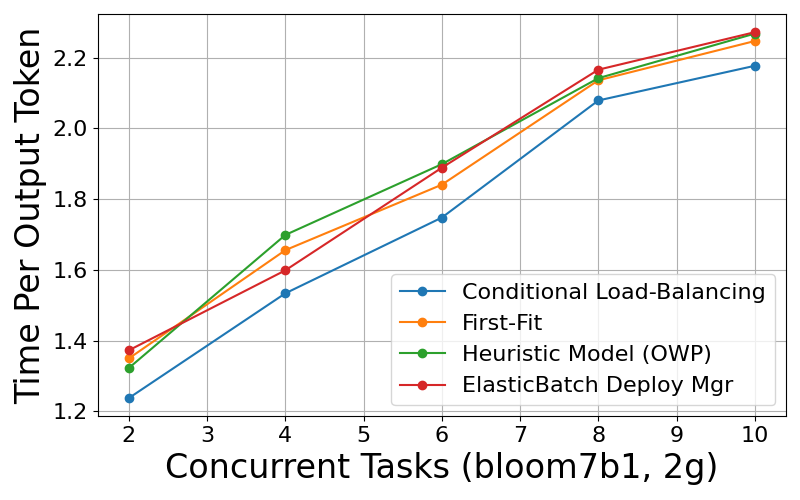}}
  \hfill
  \subfloat[Opt6.7b]{%
    \includegraphics[width=0.24\linewidth]{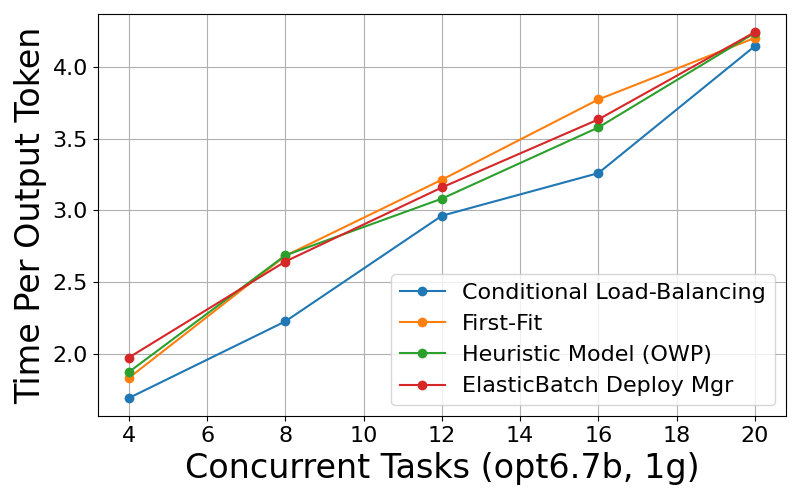}}
  \hfill
  \subfloat[Opt13b]{%
    \includegraphics[width=0.24\linewidth]{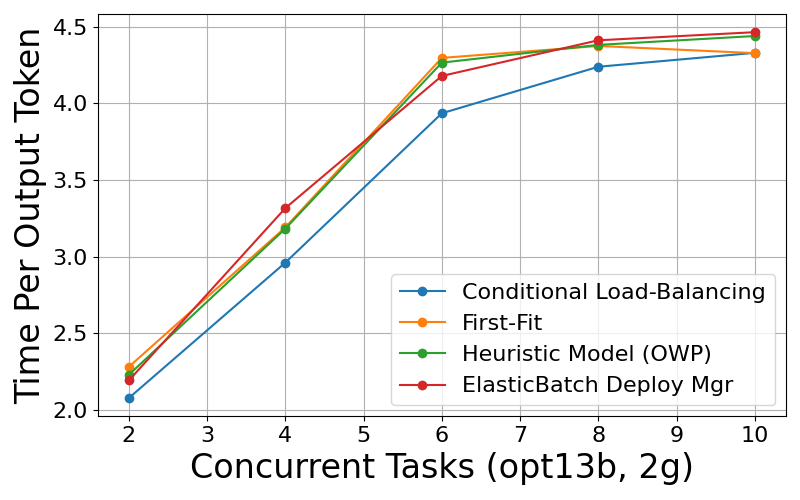}}
  \caption{Time per output token of different models under different numbers of concurrent tasks.}
  \label{fig:conditional_lb}
\end{figure*}


\subsection{Effectiveness of Conditional Load-Balancing}
We demonstrate the effectiveness of conditional load-balancing by comparing it with the naive first-fit method, the heuristic model in Optimal Workload Placement on Multi-Instance GPUs~\cite{owp}, and the deploy manager of ElasticBatch \cite{elasticbatch}. In this experiment, the load-balancing threshold is set to the average load when running all tasks on 4 GPUs. All tasks are dispatched at the first second, utilizing less than 75\% of the GPU on the node when all tasks are deployed.

Fig.~\ref{fig:conditional_lb} shows the time per output token of four models when multiple tasks are running concurrently. As the number of concurrent tasks increases, the time per output token also increases, indicating the issue of interference. Our method employs conditional load-balancing to mitigate contention, resulting in the shortest time per output token compared to the other three methods.

\subsection{Effectiveness of Dynamic Partitioning}
In this section, we evaluate the effectiveness of dynamic partitioning. We first present the runtime behavior during configuration changes, followed by a comparison against static GPU configurations, that is, GPU partitions remain the same from the start to the end of the runtime, to highlight the benefits of our approach.

Fig.~\ref{fig:dynamic-mig-cnt} shows the desired and actual number of MIG over time. Both the desired and actual numbers fluctuate. When the desired number of MIG instances increases, our method dynamically partitions the GPU to accommodate the new requirements. When the desired number decreases, our scheduler does not immediately destroy the surplus MIG instances. Instead, instances are reclaimed only when repartitioning becomes necessary for efficiency. For example, around timestamp 1900, the numbers of 2g and 4g MIG instances decrease due to an increased demand for 3g MIG instances. To fulfill this new requirement, the system initiates a dynamic repartitioning process. 

To quantitatively demonstrate the benefits of dynamic partitioning, we compare it against several static GPU configurations and select the one with the best performance to compare with. Each configuration contains the same number and size of MIG instances. The only difference between them is the placement of these instances across GPUs. The requested types of MIG instances follow the same distribution as the static configuration. Fig.~\ref{fig:dynamic} shows the wait time of four workload traces listed in Table~\ref{tab:mig-workloads}. Wait time is defined as the duration between when a task is dispatched and when it is scheduled. Static configurations result in higher wait times due to their inflexibility. Since the fragmented GPU cannot be repartitioned, a task may fail to be scheduled if the desired MIG instance is unavailable. By enabling dynamic partitioning, our scheduler consistently achieves the lowest wait time, with average improvements of at least 30\%. Dynamic partitioning offers the flexibility to adjust the GPU configuration according to the current workload, thereby significantly minimizing wait time.
\begin{figure}[tp]
    \centering
    \includegraphics[width=0.95\linewidth]{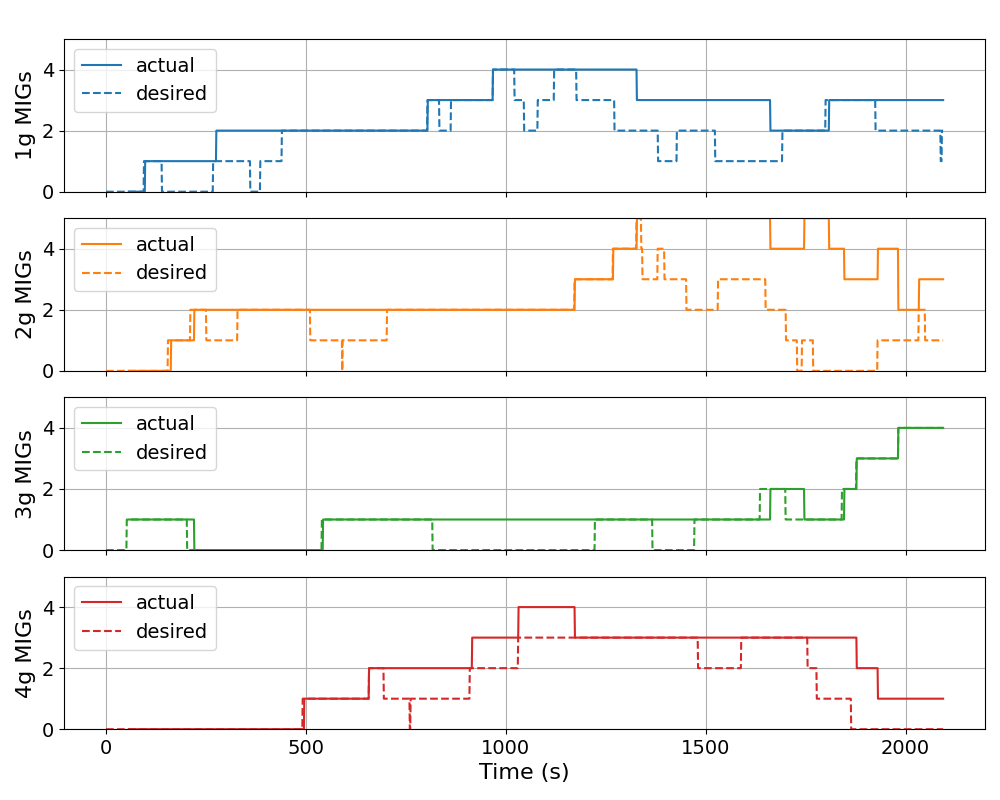}
    \caption{The desired and actual number of MIG in the system.}
    \label{fig:dynamic-mig-cnt}
\end{figure}
\begin{figure}[tp]
    \centering
    \includegraphics[width=0.95\linewidth]{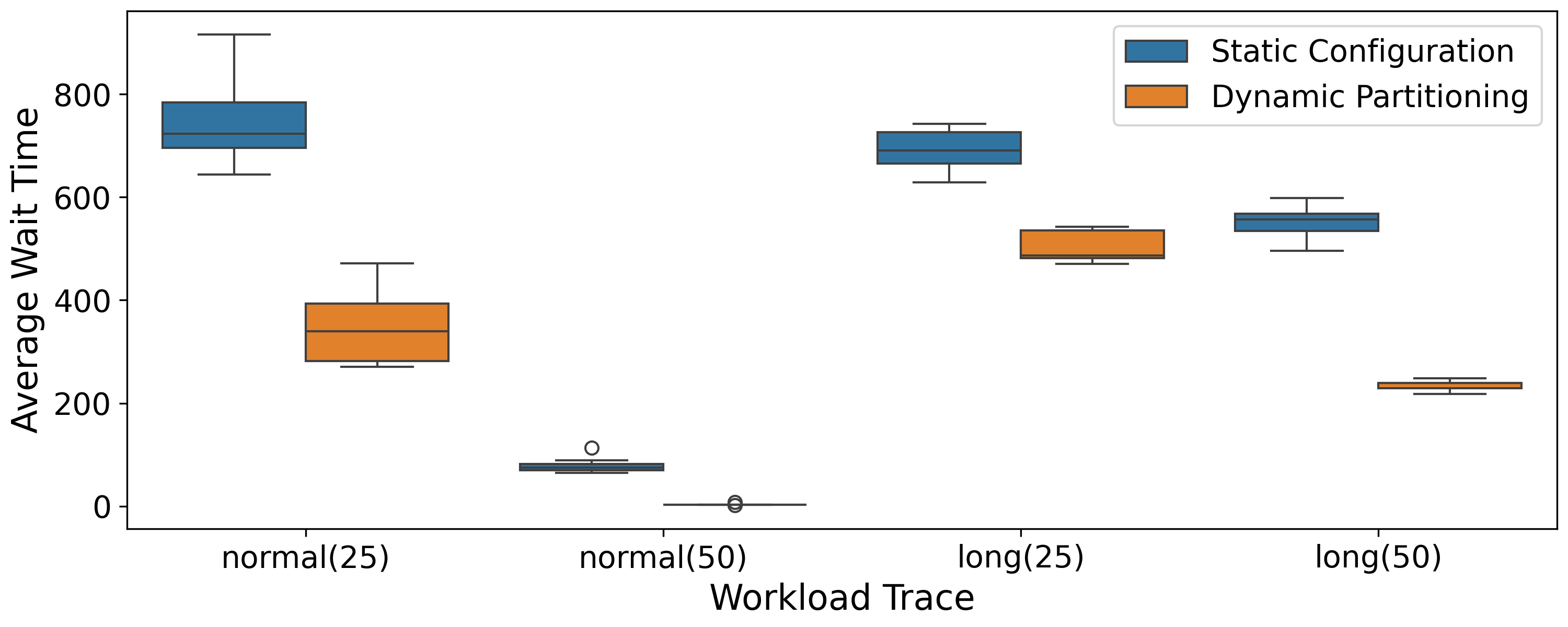}
    \caption{Average wait time of workloads under different GPU configurations.}
    \label{fig:dynamic}
\end{figure}

\subsection{Effectiveness of Job Migration}
In this section, we evaluate the effectiveness of job migration. When job migration is enabled, our scheduler determines whether to migrate a task when a task finishes and leaves the system. The advantage of job migration includes reducing GPU fragmentation and maintaining a balanced workload among GPUs. We first examine the runtime behavior of the system in terms of fragmentation and migration events. Then,  we assess the impact of job migration by measuring task execution times.

Fig.~\ref{fig:migrate-cnt} shows the system's fragmentation and corresponding migration events over time. The system's fragmentation naturally fluctuates when jobs are submitted and completed. When a job departs, the system's fragmentation may increase, which can trigger a migration event. We annotate the fragmentation peaks with migration events, demonstrating that migrations mitigate fragmentation levels.
Fig.~\ref{fig:mig-migrate} shows the average execution time of workloads when migration is either enabled or disabled. Execution is defined as the duration between when a task is scheduled and when it is completed. Enabling migration reduces execution time by more than 10\% for the normal(50) and long(50) workloads, and by more than 6\% for the normal(25) and long(25) workloads. The normal(50) and long(50) workloads have a lower arrival rate, providing more opportunities for job migration. Overall, enabling migration leads to a more balanced distribution of workloads across GPUs with less fragmentation, allowing tasks to complete faster.

\begin{figure}[tp]
    \centering
    \includegraphics[width=0.95\linewidth]{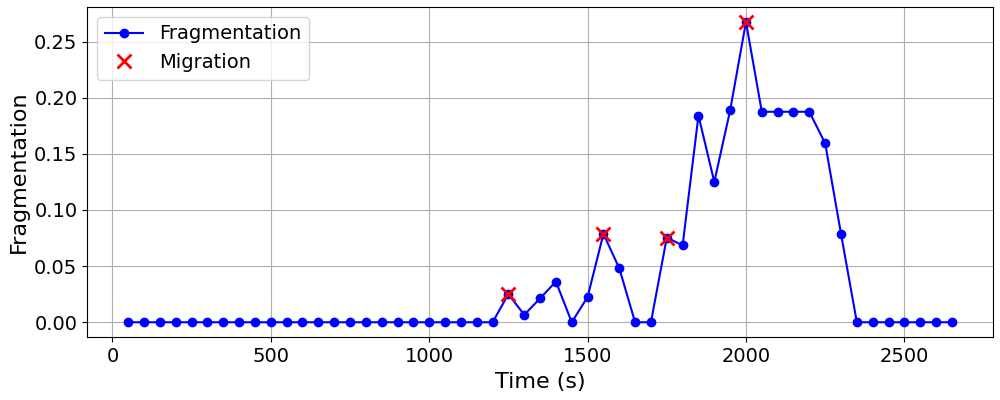}
    \caption{The fragmentation level over time. The fragmentation peaks with migration events are marked.}
    \label{fig:migrate-cnt}
\end{figure}

\begin{figure}[tp]
    \centering
    \includegraphics[width=0.95\linewidth]{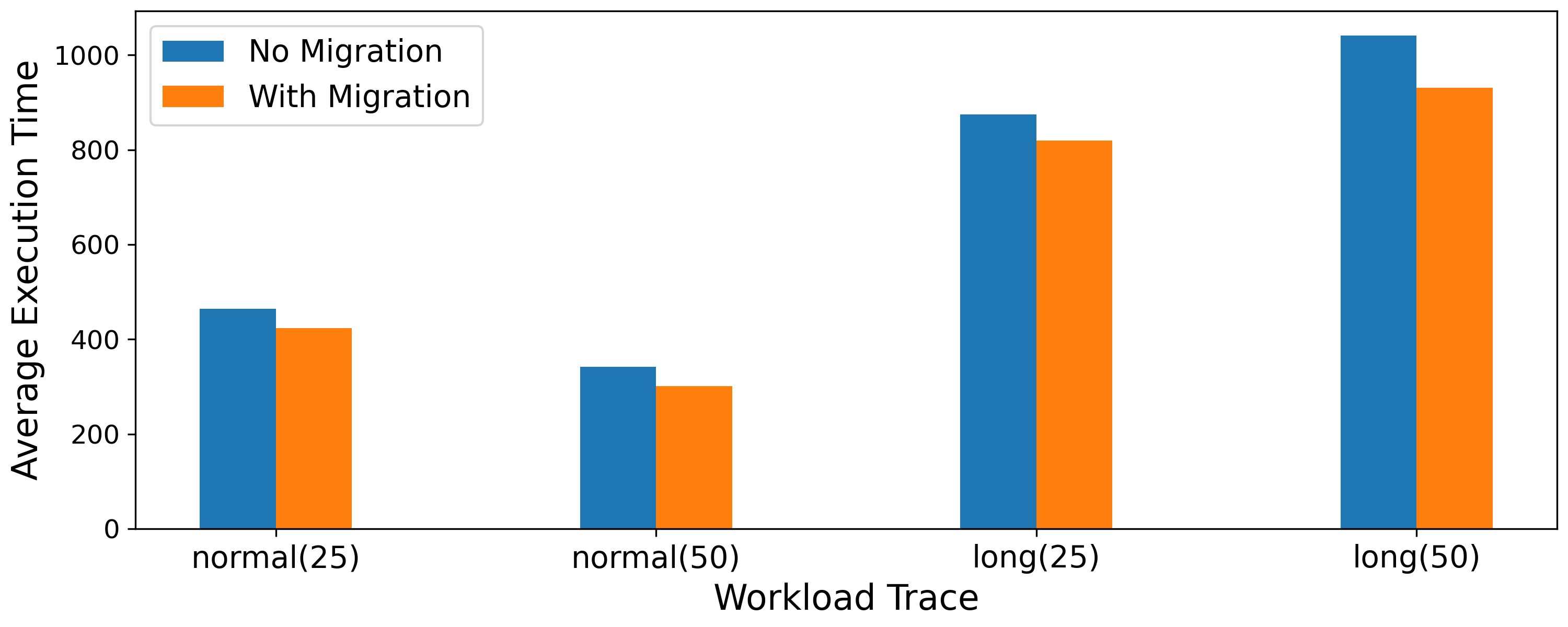}
    \caption{Average execution time of workload under job migration enabled and disabled settings.}
    \label{fig:mig-migrate}
\end{figure}

\subsection{Ablation Study}
The baseline we use in this experiment adopts first-fit scheduling with dynamic partitioning and job migration disabled. We demonstrate the advantages of enabling conditional load-balancing, dynamic partitioning, and job migration by comparing each against the baseline.
Fig.~\ref{fig:mig-ablation} shows the makespan normalized to the baseline, where makespan is defined as the sum of wait time and execution time. The results consistently show a reduction in makespan when more management features are introduced. The improvement ranges from 13\% to 35\%.

\begin{figure}[tp]
    \centering
    \includegraphics[width=0.95\linewidth]{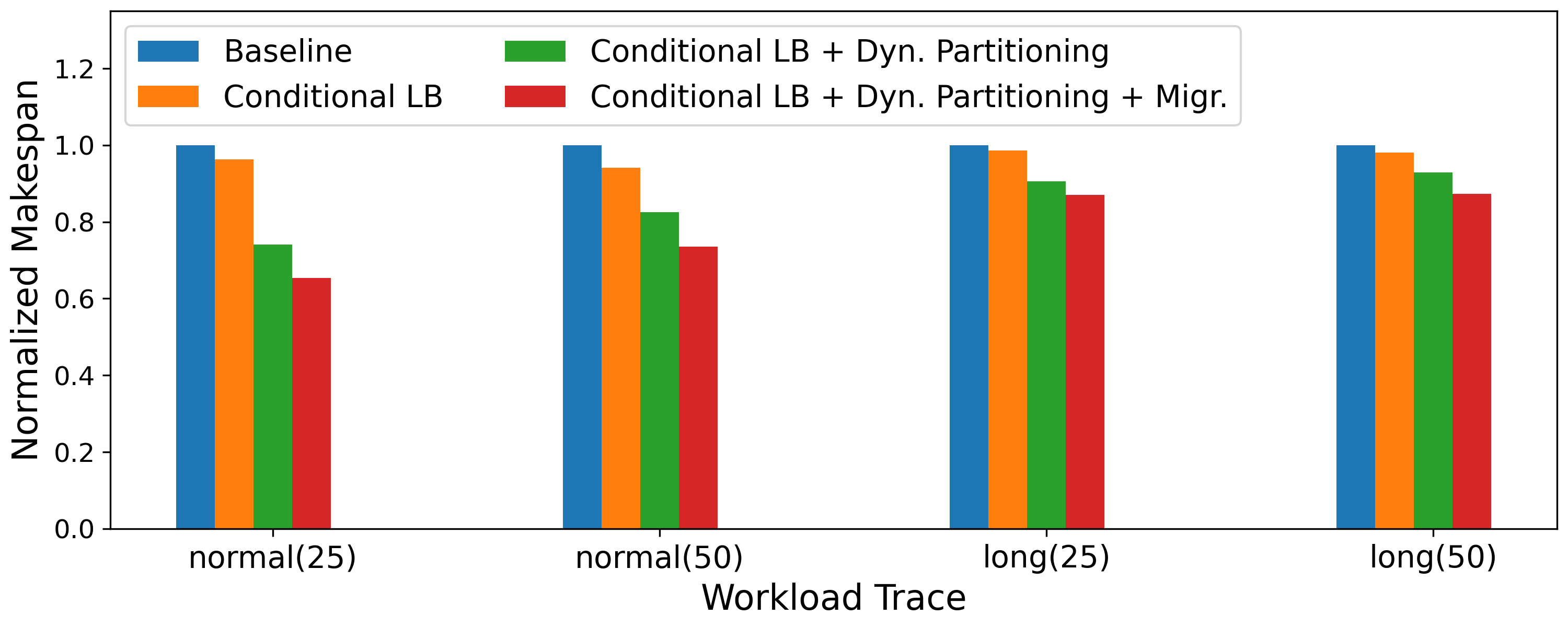}
    \caption{Makespan normalized to the baseline. LB: load balancing, Dyn.: dynamic partitioning, Migr.: migration.}
    \label{fig:mig-ablation}
\end{figure}

\section{Related Works}
\label{sec:related}

\subsection{MIG Scheduling}
Many existing works integrate the Multi-Instance GPU (MIG) feature into the scheduler due to better resource isolation and its reconfigurability. Some schedulers, like MIG-Serving~\cite{mig-serving}, ElasticBatch~\cite{elasticbatch}, ParvaGPU~\cite{parvagpu}, and FAR~\cite{FAR} are designed for offline scheduling, where the entire workload is known in advance.
MIG-Serving figures out MIG configurations for job deployment by using a heuristic greedy algorithm and a genetic algorithm.
ElasticBatch emphasizes the importance of considering dynamic inputs when scheduling with a job controller that groups jobs with similar input sizes together.
ParvaGPU blends MIG and MPS, referring to a MPS-activated MIG instance as a GPU segment. It addresses GPU partitioning challenges, including internal slack and external fragmentation by the segment configurator and the segment allocator, respectively. FAR optimizes the MIG scheduling in batches, aiming to reduce the combined makespan.

In contrast, other MIG schedulers are designed for online scheduling, where jobs arrive dynamically and have to be scheduled in real-time. MISO~\cite{miso} utilizes Multi-Process Service (MPS) to predict optimal MIG partitioning, achieving lower job completion time compared to static partitioning.
MIGRator~\cite{migrator} leverages MIG reconfigurability to support multi-tenancy continuous learning workloads, balancing SLO attainment and inference accuracy.
Paris and Elsa~\cite{paris} are elastic scheduling algorithms for reconfigurable multi-GPU inference servers. Paris reconfigures GPUs based on model and batch size, while Elsa performs heterogeneity-aware runtime scheduling. MIGER~\cite{miger} focuses on MPS atop MIG partition design, aiming to maximize the offline jobs' throughput while guaranteeing online jobs' QoS requirements. DRL~\cite{DRL} handles both online and offline cases and minimizes the combined makespan through reinforcement learning.
However, these methods overlook resource contention within the MIG instances.

Although PCIe bandwidth-aware scheduler~\cite{pcie} highlights the PCIe contention issue among MIGs, it does not actively address internal and external GPU fragmentation caused by the limited number of available GPU configurations under MIG.
Overall, prior work often underestimates resource contention within MIG and fails to address both internal and external fragmentation for online workloads. In our work, we alleviate resource contention with a conditional load-balancing algorithm, address internal fragmentation through dynamic partitioning, and handle external fragmentation by job migration. 

\subsection{GPU Multi-Tenant Systems}
There has been prior work on multi-tenant machine learning systems aimed at balancing key performance objectives such as throughput, latency, and cost. Systems like Clipper~\cite{clipper} and Nexus~\cite{nexus} focus on maximizing throughput while satisfying latency constraints. Subsequent research emphasizes improved resource utilization in addition to performance. For example, GSLICE~\cite{gslice} and gpulet~\cite{gpulet} enhance both throughput and GPU efficiency by leveraging spatial sharing mechanisms. Other systems prioritize cost efficiency. INFaaS~\cite{infaas} selects cost-effective model variants that meet user-defined accuracy and performance requirements, while iGniter~\cite{igniter} is aware of interference among co-located jobs and introduces a cost-efficient GPU provisioning strategy that ensures service-level objectives (SLOs) are met.

\section{Conclusions}
\label{sec:conc}
In this work, we leverage the capabilities of NVIDIA’s Multi-Instance GPU (MIG) architecture while addressing key challenges that are often overlooked, including resource contention and GPU fragmentation. We highlight that fragmentation under MIG differs significantly from traditional GPU sharing due to the limited number of valid configurations. To this end, we propose an online scheduling framework that integrates conditional load balancing, dynamic partitioning, and job migration. Conditional load balancing navigates the trade-off between performance and fragmentation using a tunable threshold; dynamic partitioning eliminates internal fragmentation by matching job requirements with precise MIG slices; and job migration upon departure reduces external fragmentation by reorganizing workloads to improve GPU availability. Our experimental results show that this integrated approach significantly improves system performance, reducing job wait time, execution time, and overall makespan. When all techniques are applied together, the makespan improves from 13\% to 35\%, demonstrating the practical impact of our design.

\section*{Acknowledgments}
The authors acknowledge the support from the National Science and Technology Council (NSTC) in Taiwan under grant numbers 111-2221-E-007-064-MY3.
The authors also express their appreciation for the computational resources provided by the IBM Thomas J. Watson Research Center, which were essential to this work. 

\bibliographystyle{plain}
\bibliography{ref}

\end{document}